\documentclass[showpacs,twocolumn,aps,floatfix,superscriptaddress]{revtex4}
\usepackage{amsmath,amssymb,eucal,graphicx,subfigure}

\def\l{\ell}
\def\lsim{\mathrel{\rlap{\lower4pt\hbox{\hskip1pt$\sim$}}
    \raise1pt\hbox{$<$}}}               
\def\gsim{\mathrel{\rlap{\lower4pt\hbox{\hskip1pt$\sim$}}
    \raise1pt\hbox{$>$}}}     

\begin{document}

\title{Greedy Connectivity of Geographically Embedded Graphs}

\author{Jie Sun}
\email{sunj@clarkson.edu}
\affiliation{Department of Mathematics \& Computer Science, Clarkson University, Potsdam, NY 13699-5815, USA}
\author{Daniel ben-Avraham}
\email{benavraham@clarkson.edu}
\affiliation{Physics Department, Clarkson University, Potsdam, NY 13699-5820, USA} 
\affiliation{Department of Mathematics \& Computer Science, Clarkson University, Potsdam, NY 13699-5815, USA}

\begin{abstract}
We introduce a measure of {\em greedy connectivity} for geographical networks (graphs embedded in space) and where the search for  connecting paths relies only on local information, such as a node's location and that of its neighbors.  Constraints of this type are common in everyday life applications.  Greedy connectivity accounts also for imperfect transmission across established links and is larger the higher the proportion of nodes that can be reached from other nodes with a high probability.  Greedy connectivity can be used as a criterion for optimal network design.
\end{abstract}

\maketitle

Large, complex graphs, or networks, have been the subject of much recent interest due to their ubiquity in everyday life and virtually all walks of science~\cite{reviews}.  In particular, the ability to connect any two nodes by a continuous path along the graph's edges is crucial to its function (transmitting information, controlling the spread of disease, etc.) and has been studied at length~\cite{reviews}.  

A graph $G(V,E)$ consists of a set $V$ of $N$ vertices $1,2,\dots,N$ and a set $E$
of edges, or links $(i,j)$, connecting between the nodes ($i$ and $j$).  The nodes $i$ and $j$ are {\em neighbors}.  Nodes $s$ and $t$ are {\em connected}
if a continuous path of edges $(s,v_1), (v_1,v_2),\dots,(v_{\l-1},t)$ can be found between the two nodes.  In this view,
connectivity is a {\em global} property: a complete knowledge of the graph is required to decide which pairs of nodes are connected.   

In this letter we address the question of connectivity in a different, yet commonly encountered setting.  
Consider, for example, the paradigmatic
experiment of the social psychologist Stanley Milgram~\cite{milgram}, who asked people in Omaha, Nebraska, to deliver a postcard to another person in Boston, Massachusetts.  The name and address of the target person was disclosed, but the participants were to deliver the postcards only to people they knew on a first-name basis.  If they did not know the target, the postcard was to be delivered to an acquaintance, who would then deliver it onward following the same rules, etc.  About 20\% of the cards reached their target, taking an average of $5.5$ steps, a result that gave rise to the idea of ``six degrees of separation"  and the small world phenomenon~\cite{dodds}.

Two major ingredients are different in Milgram's experiment from the usual concept of graph connectivity: (1)~Connectivity is established from {\em local} information alone --- participants knew little else beyond their own acquaintances and had no access to the full net of social contacts.  (2)~The network in question is embedded in space, i.e., each node (person) has a well defined location.  The decision who to mail the postcard to is clearly influenced by distance from the target.  This situation is not uncommon: global information is rarely available in large complex networks, while geographically embedded nets include numerous important examples, such as routers of the Internet, networks of flight connections, the electricity power grid, and neurons in the brain, to name a few, and navigation from local information is an attractive problem~\cite{chaintreau,fraig,boguna,chen,adamic,watts1,white,kleinberg1,kleinberg2,k3}.


Consider then a graph $G(V,E)$ embedded in space, and denote the geographical (Euclidean) distance between nodes by
$d(i,j)$.  We wish to establish whether a source node $s$ is connected to a target node $t$, whose location is
disclosed, relying only on local information.
Inspired by Kleinberg~\cite{kleinberg1,kleinberg2}, we model the search for connectivity by the greedy algorithm: {\em Make the next step to that neighbor that is closest to the target, provided that the distance diminishes}.  Or, symbolically, $(s,v_1,v_2,\dots,v_{\l-1},t)$ is a {\em greedy path} of length $\l$ from $s\equiv v_0$ to $t\equiv v_\l$ if for $k=1,2,\dots,\l$
\begin{equation}
\label{define}
d(v_k,t)<d(i,t),\quad {\rm and}\quad d(v_k,t)<d(v_{k-1},t),
\end{equation}
for all the neighbors $i\neq v_k$ of $v_{k-1}$.  We are assuming that the nodes are placed in a continuum so that no two pairs of nodes are at the same distance from one another.   With this understanding, greedy paths are unique. If~(\ref{define}) is fulfilled for some $\l$, we say that $s$ and $t$
are {\em greedily connected}.  

By definition, a greedy path is automatically a path.  The converse is not true.  Many other properties differentiate between connectivity and greedy connectivity:  A greedy path is not necessarily reversible --- the greedy path found from $s$ to $t$ is not always a greedy path from $t$ to $s$;  There is no transitivity --- if $i$ is greedily connected to $j$ and $j$
is greedily connected to $k$ it does {\em not} follow that $i$ is greedily connected to $k$, or in other words, the concatenation of greedy paths is not necessarily a greedy path;  If  $(s,v_1,v_2,\dots,v_{\l-1},t)$ is a greedy path
from $s$ to $t$, then $(v_i,v_{i+1},\dots,t)$ is a greedy path (from $v_i$ to $t$), however, other sub-paths  are not always greedy paths --- e.g., $(s,v_1,\dots,v_i)$ might not be a greedy path from $s$ to  $v_i$; Perhaps most surprisingly, adding links to an existing network does not necessarily increase greedy connectivity and might actually have the opposite effect.

Due to the irreversibility of greedy paths, one cannot define a greedily connected component of a graph.  Instead,
we propose measuring greedy connectivity by the quantity
\begin{equation}
GC=\sum_{s,t}\sigma_{st}e^{-\mu\l_{st}}/N(N-1)\,,
\end{equation}
where the sum runs over all $N(N-1)$ pairs of  nodes $s$ and $t$ ($s\neq t$), $\sigma_{st}=1$ if there exists a greedy path from $s$ to $t$ and is 0 otherwise, and $\l_{st}$ is the length of the greedy path
from $s$ to $t$ (when it exists).  For $\mu=0$, $GC$ is simply the fraction of all pairs that are greedily connected.  $\mu$ can be thought of as a {\em chemical potential}, or $\omega\equiv e^{-\mu}$ can be interpreted as the probability to make 
the transition across a single link successfully.  (This is important in situations such as the Milgram experiment,
where $\omega<1$.)  $GC(\omega)$ is the actual fraction of successful connections between all possible pairs of nodes
when the transmission probability across each link is $\omega$.

\begin{figure}[]
\includegraphics[width=0.45\textwidth]{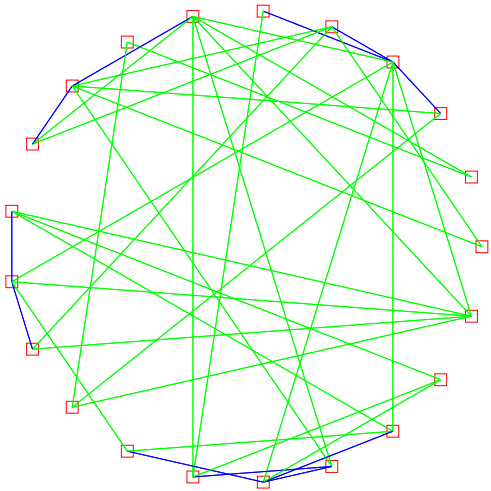}
\caption{(Color online) Circle-embedded ER graph with $N=2L+1=21$ nodes and $p=0.2$. ``Short-ranged" links, to just one or two nodes away, are highlighted in a different shade.}
\label{ER.fig}
\end{figure}

We now turn to some key examples.  Consider first an Erd\H os-R\'enyi (ER) random graph  embedded in a circle, where each link is realized with probability $p$ (Fig.~\ref{ER.fig}).  For simplicity, assume that the $N=2L+1$ nodes are equally spaced, ${\bf r}_j=(\cos\frac{2\pi j}{N},\sin\frac{2\pi j}{N})\in\mathbb{R}^2$. To avoid degeneracy of greedy paths, we introduce a small random perturbation to the location of each node.  Alternatively,
one can work with equal spacing and preserve uniqueness by making an arbitrary random choice when more than one option for a greedy step becomes available.  Distances can be measured as either $|{\bf r}_i-{\bf r}_j|$ or 
$\min\{|i-j|,N-|i-j|\}$,
to the same effect.  We opt for the latter.

Denote by $P(\l,m)$ the probability that two nodes, $m$ lattice spacings apart, are connected by a greedy path of 
$\l$ steps.  It obeys the equation
\begin{equation}
\label{PlmER}
P(\l,m)=pP(\l-1,0)+(1-q^2)\sum_{k=1}^{m-1}q^{2k-1}P(\l-1,k),
\end{equation}
($q\equiv1-p$ is the probability that a link is absent), with boundary condition $P(\l,0)=\delta_{\l,0}$.
The first term on the rhs denotes the event that there is a direct link between the target and source (probability~$p$) and
the boundary condition  tells us that the greedy path has then length 1.  The first term implied by the sum refers
to the case that the direct link is absent (prob.~$q$) but a link to at least one of the two nearest neighbors of the target exists (prob.~$1-q^2$); from there, one needs a greedy path of length $\l-1$ (since one step has already been taken) to the target at distance $k=1$, expressed by the $P(\l-1,k)$.  Successive terms model the events that increasingly more
links to the sites surrounding the target are absent.

\begin{figure}[]
\includegraphics[width=0.45\textwidth]{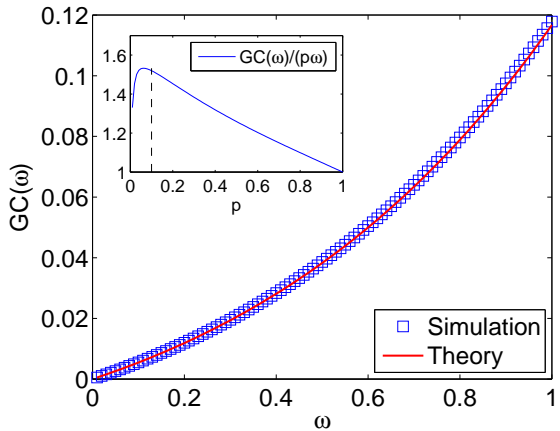}
\caption{(Color online) Greedy connectivity of circularly embedded ER graph, with $L=100$ and $p=0.05$, as a function of the transmission probability $\omega$.  Inset: The network enhancement factor, $GC(\omega)/p\omega$, as a function
of $p$, for $\omega=0.5$.}
\label{GC_ER.fig}
\end{figure}

Equation~(\ref{PlmER}) can be solved in standard ways, to yield
\begin{equation}
P_\omega(m)\equiv\sum_{\l=1}^{m}P(\l,m)\omega^{\l}=p\omega\prod_{k=2}^{m}\left[1+\omega(1-q^2)q^{2k-3}\right].
\end{equation}
Finally, using $GC(\omega)=(1/L)\sum_{m=1}^LP_{\omega}(m)$, we get
\begin{equation}
GC(\omega)=p\omega\,\frac{1}{L}\left(1+\sum_{m=2}^{L}\prod_{k=2}^{m}\left[1+\omega(1-q^2)q^{2k-3}\right]\right).
\end{equation}
It is interesting to note that $p\omega$ is the greedy connectivity that would result if the only greedy path available
between any two nodes were a direct link (that occurs with prob.~$p$).  Thus, the remaining factor is the enhancement to the $GC$ that occurs as a result of other available paths, when the direct link is absent.  This enhancement factor is bounded by $e^\omega$ and achieves its maximum near $p\sim 1/\sqrt{L}$.  Typical results for the greedy connectivity  of ER graphs, comparing our theoretical analysis to computer simulations, are shown in Fig.~\ref{GC_ER.fig}.

\begin{figure}[]
\includegraphics[width=0.25\textwidth]{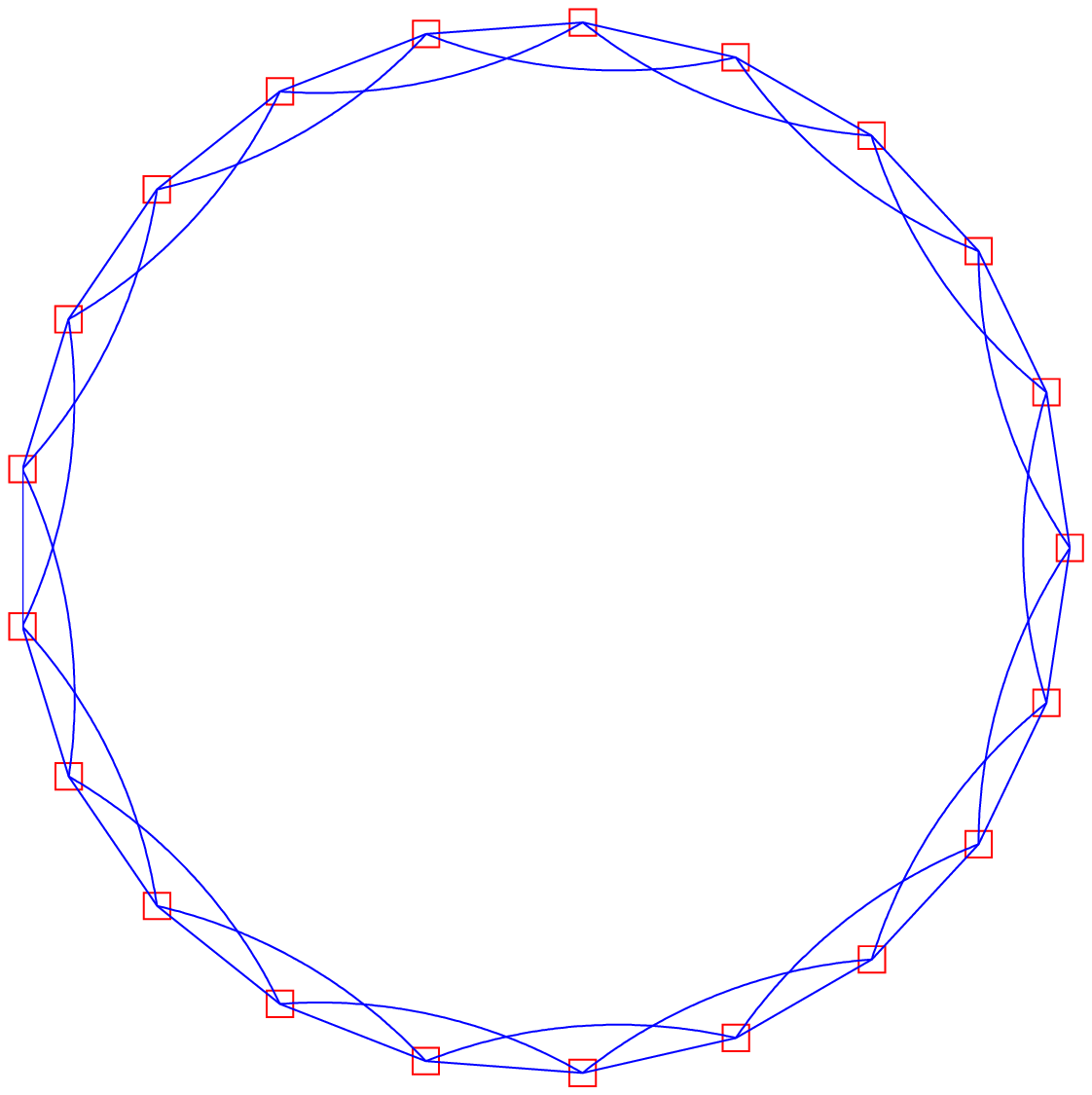}\hskip -0.15in
\includegraphics[width=0.25\textwidth]{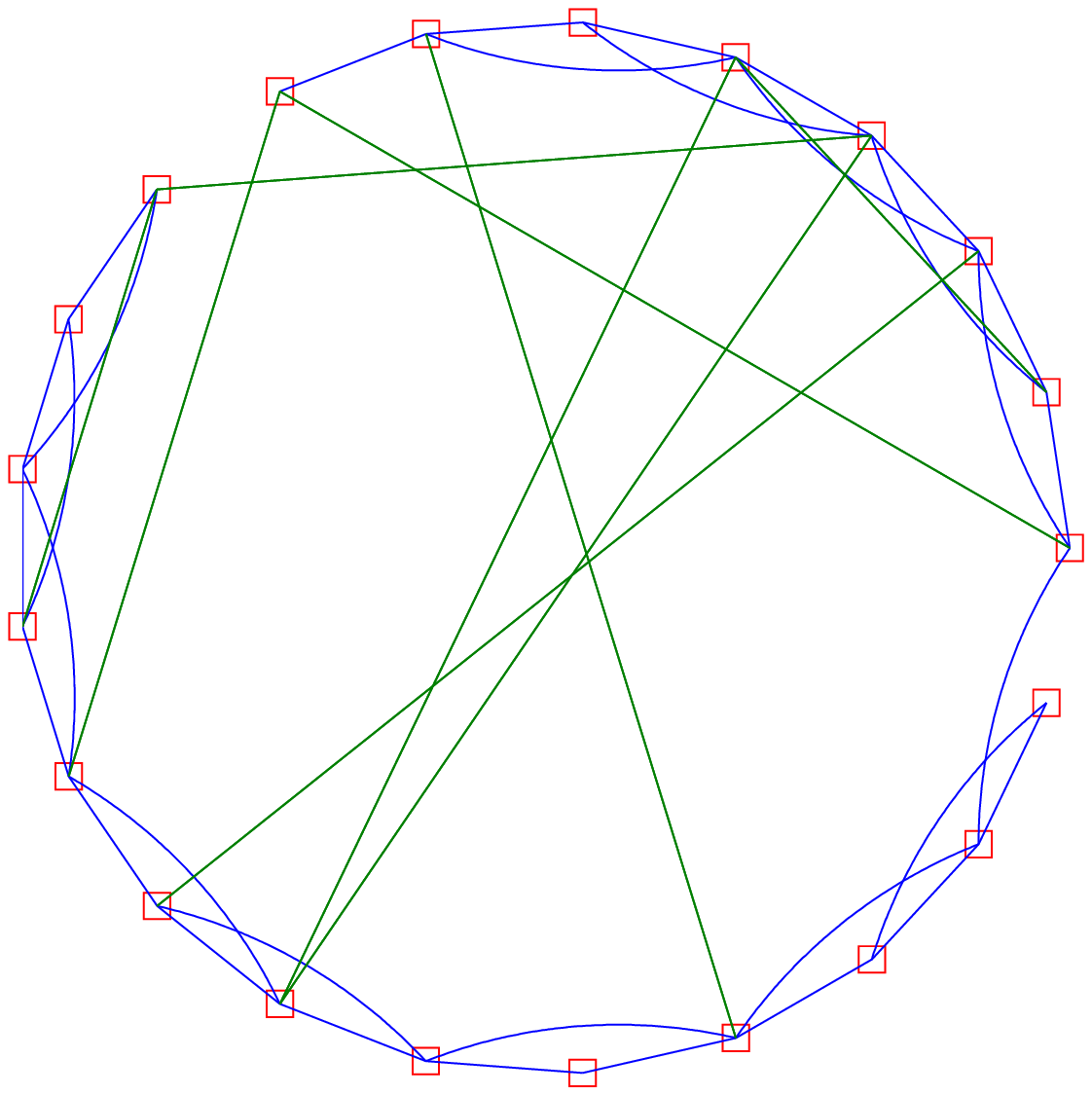}
\caption{(Color online) (a)~Circularly embedded lattice (left) with $N=2L+1=21$ nodes and $l=2$.
(b)~Circularly embedded Small-World network (right), obtained by removing a fraction $\epsilon=0.2$ of the links
and reconnecting them between random pairs of nodes.}
\label{SW.fig}
\end{figure}

Next, consider circularly embedded Small-World (SW) networks~\cite{watts}.  We start with the underlying ``lattice" configuration, where each of the $N=2L+1$ nodes is connected to $l$-nearest neighbors on either side (Fig.~\ref{SW.fig}a).   As before, the nodes are slightly perturbed from their lattice centers, to avoid degeneracy of greedy paths.
The equation for $P(\l,m)$ reads
\begin{equation}
P(\l,m)=\delta_{\l,  \lceil m/l \rceil},
\end{equation}
where $\lceil x \rceil$ is the smallest integer greater or equal to $x$.  We then have $P_{\omega}(m)=\sum_\l P(\l,m)\omega^{\l}=\omega^{\lceil m/l \rceil}$, and
\begin{equation}
\begin{split}
&GC(\omega)=\frac{1}{L}\sum_{m=1}^{L}P_{\omega}(m)\cr
&=\frac{l}{L}\left(\omega+\omega^2+\cdots+\omega^{L/l}\right)=p\omega\,\frac{1-\omega^{1/p}}{1-\omega},
\end{split}
\end{equation}
where, for simplicity, we have assumed that $L$ is a multiple of $l$, and we write $l/L=p$ for comparison
with ER graphs (this yields the same number of links in either case).  Indeed, it is interesting to note that 
the greedy connectivity of the lattice is always larger than that of an equivalent ER graph.  The lattice architecture
guarantees that any two sites are connected, yet the  typical distances
are order $N$, rather than $\ln N$, as in ER graphs.  The benefits seem to get the upper hand.

To achieve the small-world  effect, a fraction $\epsilon$ of the links are removed and are then reconnected between randomly selected pairs of nodes (but avoiding multiple connections between any pair), see  Fig.~\ref{SW.fig}b.  Even a small
fraction $\epsilon$ of randomly rerouted links reduces the typical shortest path between nodes, from $O(N)$ to $O(\ln N)$.
We now show that the fraction $\epsilon$ can be optimized to attain a maximum in the greedy connectivity (in particular, outperforming the lattice, for which $\epsilon=0$).

The equation for the $P(\l,m)$ of SW networks is
\begin{equation}
\label{PlmSW}
\begin{split}
&P(\l,m)=pP(\l-1,0)+(1-q^2)\sum_{k=1}^{m-3}q^{2k-1}P(\l-1,k)\cr
&+(1-q'q)q^{2m-5}\{P(\l-1,m-2)
+q'qP(\l-1,m-1)\},
\end{split}
\end{equation}
where we have specialized to the case of $l=2$.  For links spanning nodes more than $l=2$ lattice spacings apart
the equation is the same as for ER graphs, with $p\equiv\epsilon l/L$, now the effective probability of random long-range links.  The only difference is when the first greedy step is to a site within $l$ spacings; these require $l$ specialized terms (the last two terms, in our case) because the probability of such short-range links is $p'\equiv1-\epsilon+p$ (and $q'=1-p'$), rather than $p$.
Eq.~(\ref{PlmSW}) is valid for $m\geq3$. The boundary conditions are revised, for the very same reason:
\[
P(\l,1)=\delta_{\l,1}p',\quad
P(\l,2)=\delta_{\l,1}p'+\delta_{\l.2}(1-q'q)p'q'\,.
\]

\begin{figure}[]
\includegraphics[width=0.45\textwidth]{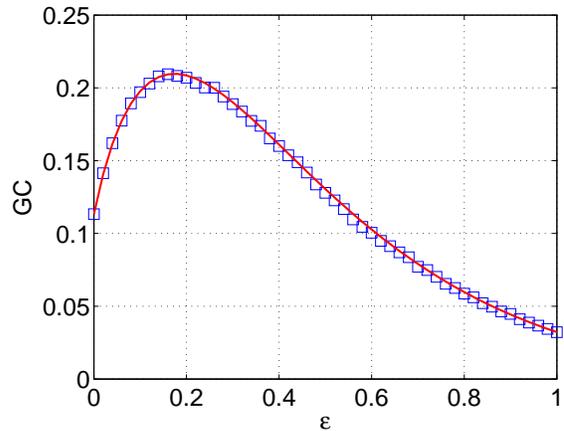}
\caption{(Color online) Greedy connectivity of circularly embedded SW graph, with $L=100$, $l=2$, and $\omega=0.85$, as a function of the fraction of random links, $\epsilon$.  Theoretical results (solid curve) are compared to numerical simulations (symbols).}
\label{GC_SW.fig}
\end{figure}

Eq.~(\ref{PlmSW}) can be solved by standard techniques.  The final expression we obtain for $GC(\omega)$ is
too cumbersome to list here, but it agrees perfectly well with numerical simulations, as shown in Fig.~(\ref{GC_SW.fig})
for one typical case.  Note the maximum in $GC$, about $\epsilon\approx0.2$, which is nearly twice as large as the $GC$ of the corresponding lattice, at $\epsilon=0$, and about 7 times as large as the corresponding ER network, at $\epsilon=1$.  Qualitatively similar results are obtained for most other link densities, $l\geq 2$.  For $l=1$, the maximum $GC(\omega)$ occurs always at $\epsilon=0$, that is, for the underlying lattice (a simple ring).

Our third and last example is that of circularly embedded scale-free (SF) networks.  As usual, the $N=2L+1$ nodes are to be placed on a ring, slightly perturbed from their lattice locations, and we start with a single node at $(0,0)$.  We then construct a scale-free net according to the {\em redirection} algorithm of Krapivsky and Redner~\cite{KR}: Each new node is brought in (to a random location) and is connected to one of the existing nodes, selected randomly, with probability $1-r$.  With probability $r$, the connection is redirected to the {\em ancestor} of that node (the node it was attached to first, when it was added to the net).  This yields a graph
with scale-free degree distribution, $P(k)\sim k^{-\gamma}$, $\gamma=1+1/r$.  Because the networks built in this way are actually trees, the average degree is $\langle k\rangle=2$, so the procedure has the advantage of keeping a constant
density of links even as $r$, or $\gamma$, is varied.

\begin{figure}[]
\includegraphics[width=0.49\textwidth]{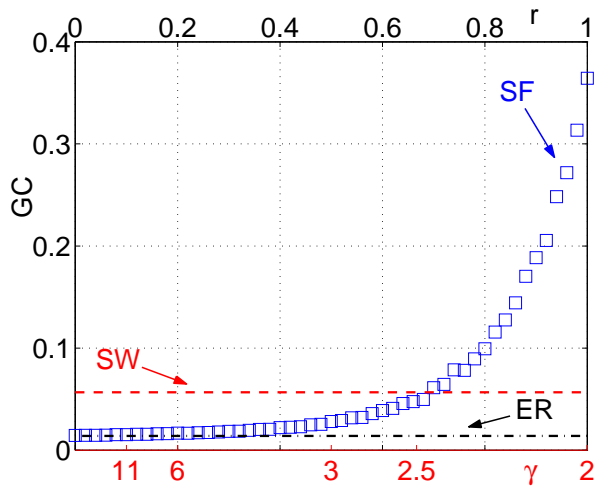}
\caption{(Color online) Greedy connectivity of circularly embedded SF nets (with $L=100$ and $\omega=0.85$) as a function of their degree distribution exponent, $\gamma=1+1/r$.  The $GC$ of equivalent ER graphs (-- $\cdot$ --) and
SW graphs (-- -- --) is shown for comparison.}
\label{GC_SF.fig}
\end{figure}

In Fig.~\ref{GC_SF.fig} we present data culled from computer simulations of circularly-embedded SF nets.  In the limit of $r\to0$ ($\gamma\to\infty$) the networks are trees with a {\em narrow} degree distribution, similar to ER graphs.  The $GC$ in that limit is equal to that of equivalent ER graphs (with the same link density).  In the opposite limit of $r\to1$ ($\gamma\to2$) all the links are redirected to one ``super-hub" and we get a {\em star graph}.  It is easy to show that in this case $GC=(1/2)\omega^2$ (for $L\gg1$).
As $\gamma$ decreases from $\infty$ to $2$ the GC of the SF networks increases monotonically, the largest increase occurring between $3\gsim\gamma>2$, corresponding to the regime encountered in most frequent applications~\cite{reviews}.  Note that SF networks with $\gamma\lsim 2.5$ exhibit a greater $GC$ than that of the optimal corresponding SW net
(in this case, of $l=1$, a simple ring, or $\epsilon=0$).

In summary, we have introduced a measure of {\em greedy} connectivity for geographical networks (graphs embedded in space) and where the search for a connecting path might rely only on local information, such as a node's location and that of its neighbors (the ones linked to it).  This is useful in a host of situations where the networks are large and complex
and global information is not available, or relying on it is impractical due to the network's size.  Greedy connectivity  is larger the larger the fraction of connected nodes.

Greedy connectivity generalizes the Kleinberg navigation problem (by which it is inspired) in several ways, most importantly, in that nothing is presumed about the network structure; the existence of a greedy path between any two nodes is {\em not}  required, and the probability of transmission across any given
link, $\omega$, now plays a defining role.  Indeed, Kleinberg-like greedy paths, of minimal length, can be found for any geographically embedded network by maximizing $GC(\omega)$ in the limit of $\omega\to 0$ (or $\mu\to\infty$).

An important feature, suggested by the examples analyzed herein, is that greedy connectivity can be enhanced and optimized by varying the network architecture, including the geographical placement of the nodes.  This is perhaps the richest venue for future applications.

\acknowledgments

We thank Takashi Nishikawa for many useful discussions. We are grateful to the NSF, award PHY-0555312 (DbA), for partial support of this work.


\begin{thebibliography}{99}

\bibitem{reviews} R.~Albert and A.-L.~Barab\'asi, {\em Rev. Mod. Phys.} {\bf74}, 47 (2002);
M.~E.~J. Newman, {\em SIAM Rev.} {\bf 45}, 167 (2003);
M.~E.~J. Newman, A.-L.~Barab\'asi, and D.~J.~Watts, {\em The Structure and Dynamics of Networks} (Princeton University Press, 2006);
A.~Barrat, M.~Barth\'elemy, and A.~Vespignani, {\em Dynamical Processes on Complex Networks} (Cambridge, 2008).


\bibitem{milgram}
S.~Milgram,
{\em Psych. Today}, {\bf2}, 60--67 (1967).


\bibitem{dodds}
For a more recent experimental study, involving emails, see: P.~S.~Dodds, R.~Muhamad, and D.~J.~Watts,
{\em Science} {\bf301}, 827 (2003).

\bibitem{chaintreau}
A.~Chaintreau, P.~Fraigniaud, and E.~Lebhar,
in {\em Automata, Languages and Programming (LNCS)} {\bf5125}, 133 (Springer, 2008).

\bibitem{fraig}
P.~Fraigniaud and C.~Gavoille,
in {\em Proceedings of the 20th Symposium on Parallelism in Algorithm and Architectures}, pp 62--69 (ACM, New York, NY, 2008).


\bibitem{boguna}
M.~Bogu\~n\'a, D.~Krioukov, and K.~C.~Claffy, 
{\em Nature Physics}, doi:10.1038/nphys1130 (2008).


%


\bibitem{chen} J.-Z.~Chen, W.~Liu, and J.-Y.~Zhu, 
{\em Phys. Rev. E} {\bf73}, 056111 (2006).


\bibitem{adamic} L.~A.~Adamic and E.~Adar, 
{\em Social Networks} {\bf 27}, 187--203 (2005);
L.~A.~Adamic, R.~M.~Lukose, A.~R.~Puniyani, and B.~A.~Huberman, 
{\em Phys. Rev. E} {\bf64}, 046135 (2001).

\bibitem{watts1} D.~J.~Watts, P~ S.~Dodds, and M.~E.~J.~Newman, 
{\em Science} {\bf296}, 1302 (2001).

\bibitem{white}
H.~C.~White,
{\it Social Forces} {\bf49}, 259 (1970).


\bibitem{kleinberg1}
J.~Kleinberg,
\newblock {\em Nature} {\bf406}, 845 (2000).

\bibitem{kleinberg2}
J.~Kleinberg,
\newblock In {\em Proc. 32nd ACM Symp. on Theory of
  Computing}, 163--170 (2000).
  
\bibitem{k3}  
M.~R.~Roberson and D.~{ben-Avraham},
{\em Phys. Rev. E} {\bf74}, 017101 (2006);
S.~Carmi, S.~Carter, J.~Sun, and D.~ben-Avraham, {\em Phys. Rev. Lett.} {\bf102}, 238702 (2009);
C.~C.~Cartozo and P.~De~Los~Rios, 238703 (2009).
  
\bibitem{watts}
D.~J.~Watts and S.~H.~Strogatz, {\em Nature} {\bf393}, 440 
(1998);
D.~J.~Watts, {\em Small Worlds: The Dynamics of Networks between Order and Randomness} (Princeton University Press, Princeton, New Jersey, 1999).  
\bibitem{KR} P.L. Krapivsky and S. Redner, {\it Phys. Rev. E} {\bf63}, 066123 (2001);
{\it J. Phys. A} {\bf35}, 9517 (2002); J. Kim, P.L. Krapivsky, B. Kahng, and S. Redner, {\it Phys. Rev. E} {\bf66}, 055101(R) (2002); H.~D.~Rozenfeld and D.~ben-Avraham, {\em Phys. Rev. E} {\bf70}, 056107 (2004).
  



\end{thebibliography}
\end{document}